\documentclass[10pt]{iopart}
\usepackage{graphicx}
\usepackage{color}

\input epsf.sty
\begin{document}

\title[Characterization of relaxation processes in interacting vortex matter]{Characterization of relaxation processes in interacting vortex matter 
       through a time-dependent correlation length}
\author{Michel Pleimling and Uwe C. T\"{a}uber}
\address{Department of Physics, Virginia Tech, Blacksburg, VA 24061-0435, USA}
\date{\today}


\begin{abstract}
Vortex lines in type-II superconductors display complicated relaxation processes
due to the intricate competition between their mutual repulsive interactions and
pinning to attractive point or extended defects. We perform extensive Monte 
Carlo simulations for an interacting elastic line model with either point-like 
or columnar pinning centers. From measurements of the space- and time-dependent
height-height correlation function for lateral flux line fluctuations, we 
extract a characteristic correlation length that we use to investigate different
non-equilibrium relaxation regimes. The specific time dependence of this
correlation length for different disorder configurations displays characteristic 
features that provide a novel diagnostic tool to distinguish between point-like 
pinning centers and extended columnar defects.
\end{abstract}


\maketitle

\section{Introduction}

The characterization and optimization of the properties of vortex matter in 
disordered type-II superconductors has been a prominent research focus in 
condensed matter physics over the past two decades \cite{Blatter_etal:94,
Nattermann_Scheidl:00}. Detailed experimental and theoretical studies of 
interacting magnetic flux lines subject to different types of attractive
pinning centers have yielded rich equilibrium phase diagrams with novel glassy 
states as well as an intriguing complexity in the associated dynamical 
phenomena. The mean-field Abrikosov flux lattice in pure type-II 
superconductors is already destroyed at low temperatures by point-like defects 
and replaced by a disorder-dominated vortex glass phase, wherein the vortices 
are collectively pinned, and display neither translational nor orientational 
long-range order \cite{FisherM:89, Feigelman_etal:89, Nattermann:90,
FisherD_FisherM_Huse:91}. Firm theoretical arguments furthermore support the 
existence of a topologically ordered dislocation-free Bragg glass phase at low 
magnetic fields or sufficiently weak disorder \cite{Giamarchi_LeDoussal:94,
Giamarchi_LeDoussal:95, Kierfeld_Nattermann_Hwa:97, Fisher:97, 
Giamarchi_LeDoussal:97, Nattermann_Scheidl:00}. Extended attractive defects 
such as columnar pins localize the flux lines in their vicinity; the resulting 
low-temperature Bose glass phase is distinguished from the vortex or Bragg 
glass through its infinite tilt modulus or vanishing linear response to a 
magnetic field rotation (transverse Meissner effect) \cite{Nelson_Vinokur:92, 
Lyuksyutov:92, Nelson_Vinokur:93, UCT_Nelson:97}. It is important to note that
transverse flux line fluctuations become strongly suppressed in the Bose glass, 
in stark contrast to the disorder-induced line roughening in the presence of 
randomly distributed point pinning centers.

The quite distinct flux line fluctuation spectra that result for various
pinning centers should naturally be observable in non-equilibrium relaxation
processes, wherein the vortex system slowly approaches a low-temperature 
equilibrium glassy state after being initialized in a very different
configuration. Observing these relaxation features could thus provide an
effective characterization tool for type-II superconducting samples. Previous 
numerical studies of the relaxation processes that take place in the glassy 
phases of vortex matter focused on a variety of one- and two-time quantities, 
as for example the radius of gyration, the two-time vortex line height-height 
autocorrelation function, or the two-times vortex line roughness, to name but 
a few \cite{BCD1, BCD2, BCI, IBKC, Pleimling2011, Dobramysl2013, Assi2015}.
From these investigations the following coherent picture is emerging.

First, one notes that the non-equilibrium properties deep inside the glassy 
phase are very different if both attractive and repulsive defects are 
considered \cite{BCD1,BCD2} or if only attractive pinning centers are present 
\cite{Pleimling2011, Dobramysl2013, Assi2015}.
For vortices in type-II superconductors, however, material defects are 
exclusively attractive, thus providing a pinning landscape with deep wells in 
an otherwise flat potential. This is the case that we consider in this paper.

Second, relaxation processes for vortex matter in type-II superconductors are 
in general dominated by two strongly competing effects: the pinning of the flux 
lines due to attractive defects and the repulsive interaction between vortices. 
In the presence of thermal fluctuations various subtle crossover scenarios are 
encountered. For example, for defects of intermediate strength a two-step 
relaxation, reminiscent of that observed in structural glasses, is revealed in 
the two-time height-height autocorrelation function \cite{Pleimling2011, 
Dobramysl2013, Assi2015}. Furthermore, the flux line relaxation characteristics 
are found to markedly depend on the type of disorder present in the sample 
\cite{Dobramysl2013, Assi2015}, and quite complex non-universal features emerge
that vary with intrinsic material as well as external control parameters such 
as the temperature and the external magnetic field that sets the overall vortex 
density.

Important insights into relaxation properties of systems with slow dynamics can
also be gained from the investigation of time- and space-dependent correlation
functions and from the time dependence of characteristic growing length scales 
derived from these quantities; see Refs.~\cite{Kis96,Fra00,Cor10,Par10, Cor11, 
Cor12, Par12, Cor13, Man14} for various recent examples.
 In 
Refs.~\cite{Schehr_LeDoussal_04,Schehr_Rieger_05} the time-dependent 
length 
scale for a two-dimensional array of
 vortices in the presence of weak point-like 
disorder was calculated
 analytically and the resulting aging properties were 
studied. 
In this present work, we employ extensive Monte Carlo simulations to 
investigate non-equilibrium relaxation processes for interacting flux lines in 
disordered type-II superconductors through a time-dependent characteristic 
length distilled from the space-time transverse vortex height-height correlation
function. Using an effective description of the vortices as thin fluctuating 
elastic lines, valid in the extreme London limit and low flux density, we study 
systematically how the different contributions to the Hamiltonian affect this 
growing correlation length in distinct temporal regimes. We specifically compare 
two different types of attractive pinning centers, namely randomly distributed 
point-like disorder and correlated columnar defects, and identify typical 
signatures in the associated correlation lengths for these different material 
defect classes, as the vortex system equilibrates towards the disorder-dominated
Bragg or Bose glass phase, respectively. 

The paper is organized in the following way: In the next Section we recall the 
effective Hamiltonian pertaining to the elastic line description of interacting 
magnetic vortices and discuss the model parameters used in our numerical 
simulations. We then introduce the space- and time-dependent transverse 
height-height correlation function and the time-dependent characteristic length 
derived from that quantity. Section~III contains our results, both for 
point-like as well as for correlated defects. We conclude in Section~IV with a 
brief summary and outlook.

\section{Model and observables}

In our study we consider three-dimensional vortex systems composed of $N$ 
interacting flux lines. In the extreme London limit, where the superconducting
coherence length is assumed much smaller than the London penetration depth,
vortex lines in type-II superconductors are well described by a 
coarse-grained elastic line model \cite{Nelson_Vinokur:93}. Introducing the 
trajectory ${\bf r}_j(z) = x_j(z) {\bf \hat{x}} + y_j(z) {\bf \hat{y}}$ of flux
line $j$, where $z$ represents the coordinate in the direction of the applied 
magnetic field perpendicular to the $xy$ plane, the effective Hamiltonian for a 
sample of thickness $L$ is given by
\begin{eqnarray}
   H_N &=& \frac{\tilde{\epsilon}_1}{2} \sum_{j=1}^N \int_0^L \bigg\arrowvert
   \frac{d{\bf r}_j(z)}{dz} \bigg\arrowvert^2 \, dz 
   + \sum_{j=1}^N \int_0^L V_D\bigl( {\bf r}_j(z) \bigr) \, dz \nonumber \\
   && + \frac{1}{2} \sum_{i \ne j} \int_0^L V\bigl(|{\bf r}_i(z)-{\bf r}_j(z)| 
   \bigr) \, dz \ .
\label{hamilt}
\end{eqnarray}
The first term here describes the elastic line tension, with the tilt modulus 
$\tilde{\epsilon}_1 = 0.18\,\epsilon_0$ for YBCO, where 
$\epsilon_0 = (\phi_0 / 4 \pi \lambda_{ab})^2$, with the magnetic flux quantum 
$\phi_0 = h c / 2 e$ and the London penetration depth $\lambda_{ab}$, sets the 
interaction energy scale. Using for our simulation parameter values typical for 
YBCO, we take $\lambda_{ab} = 34\,b_0$ with $b_0 = 35$ \AA~and thus
$\epsilon_0 \approx 1.9 \times 10^{-6}$ erg/cm \cite{Blatter_etal:94}. The 
second contribution in (\ref{hamilt}) captures the disorder-induced attractive 
potential due to localized pinning centers, which we model as potential wells 
with in-plane radius $b_0$ and depth $U_0 = p \, \epsilon_0$, with the strength
parameter $p$ varying between $0$ and $0.20$. The third term finally represents
the repulsive vortex-vortex pair interaction. Consistent with the extreme 
London limit, this repulsion is purely in-plane between flux line elements and 
given by $V(r) = 2 \epsilon_0 K_0 (r/\lambda_{ab})$, where $K_0$ denotes the 
zeroth-order modified Bessel function. As $V(r)$ decreases exponentially for 
$r \gg \lambda_{ab}$, we truncate this vortex interaction at half of the 
lateral system size \cite{Pleimling2011}.

Using a discretized version of the Hamiltonian (\ref{hamilt}) and periodic
boundary conditions, we perform standard Metropolis Monte Carlo simulations 
\cite{Das2003, Bullard2008} at temperature $T = 10$ K, corresponding to 
$k_B T / b_0 \epsilon_0 \approx 0.002$. This temperature choice assures that 
the system resides deep in a disorder-dominated glassy regime, namely most 
likely the Bragg glass phase for the case of point pinning centers, and the 
Bose glass for extended columnar pins \cite{Pleimling2011}. Spatial 
discretization is done in the $z$ direction in such a way that we have $L$ 
equidistant layers separated by the basic microscopic distance scale $b_0$. 
Our system contains $N = 16$ vortices, with the number of flux line elements 
per vortex ranging from $L = 640$ to $L = 2560$. The in-plane dimensions are
set to $L_x = \frac{2}{\sqrt{3}} \times 8 \lambda_{ab}$ and 
$L_y = 8 \lambda_{ab}$, and we checked that the vortices arrange in a 
triangular Abrikosov flux lattice in the absence of disorder. In each layer, 
we place $N_D = 1116$ pinning centers. For the case of random point defects, 
these pinning centers are randomly distributed and chosen independently for
each layer. For columnar defects aligned parallel to the magnetic field along
the $z$ direction, we instead repeat the same spatial distribution pattern for 
each layer. As usual, we define as a Monte Carlo time step a consecutive series
of $N L$ updates; {\em i.e.}, on average every flux line element is updated 
once during each Monte Carlo time step. In the following, all distances and
length scales are measured in units of $b_0$, and time in terms of Monte Carlo
steps.

In order to study the build-up of spatial correlations, we prepare an 
out-of-equilibrium initial state by randomly placing straight lines, oriented 
along the $z$ direction, in the system. While this initial configuration is 
well-suited for our investigation, it is not a state that can be set up easily
in experiments. Initial out-of-equilibrium conditions closer to experimentally 
realizable situations can be achieved through temperature or field quenches, as 
discussed recently in Ref.~\cite{Assi2015}. 

After having prepared the system in this initial state, we bring the system in 
contact with a heat bath at temperature $T = 10$ K and follow its subsequent
time evolution by measuring the space- and time-dependent height-height 
correlation function for lateral flux line displacements,
\begin{equation} \label{eq:C}
  C(r,t) = \frac{1}{2} \bigl[ C_x(r,t) + C_y(r,t) \bigr] \ ,
\end{equation}
with
\begin{equation}
  C_x(r,t) = \Big\langle \bigl[ x_j(z,t) - \overline{x}_j(t) \bigr] \, 
  \bigl[ x_j(z+r,t) - \overline{x}_j(t) \bigr] \Big\rangle \ ,
\end{equation}
and similarly for $C_y(r,t)$. Here $x_j(z,t)$ is the $x$ coordinate of the flux
line element $z$ of line $j$ at time $t$, whereas $\overline{x}_j(t)$ 
represents the mean $x$ position of line $j$. The height-height correlation 
results from a fourfold average: an average over all $N \, L$ vortex elements, 
over several noise realizations, over different initial configurations, and 
over independent disorder realizations. The data discussed in the following 
result from averaging over typically 40 independent runs for non-interacting 
vortex lines, whereas for interacting lines we averaged over at least 400 
different runs.

A time-dependent correlation length $\xi(t)$ can be extracted from the 
height-height correlation function (\ref{eq:C}) by imposing the condition that 
for a fixed value of $t$ the correlation function $C(r,t)$ has decreased by a 
factor $C_0$ from its maximum value $C(0,t)$:
\begin{equation} 
\label{eq:C2}
  C\bigl( \xi(t),t \bigr) / C(0,t) = C_0 \ .
\end{equation}
We have used $C_0 = 0.5$ in the following, but have verified that the temporal 
evolution of the characteristic length $\xi(t)$ is qualitatively unchanged when
a different value is chosen. An alternative length scale can be obtained from 
integral estimators \cite{Belleti2009}. Comparing these two approaches we did 
not find any significant qualitative differences.

\begin{figure}[h]
\includegraphics[width=0.85\columnwidth]{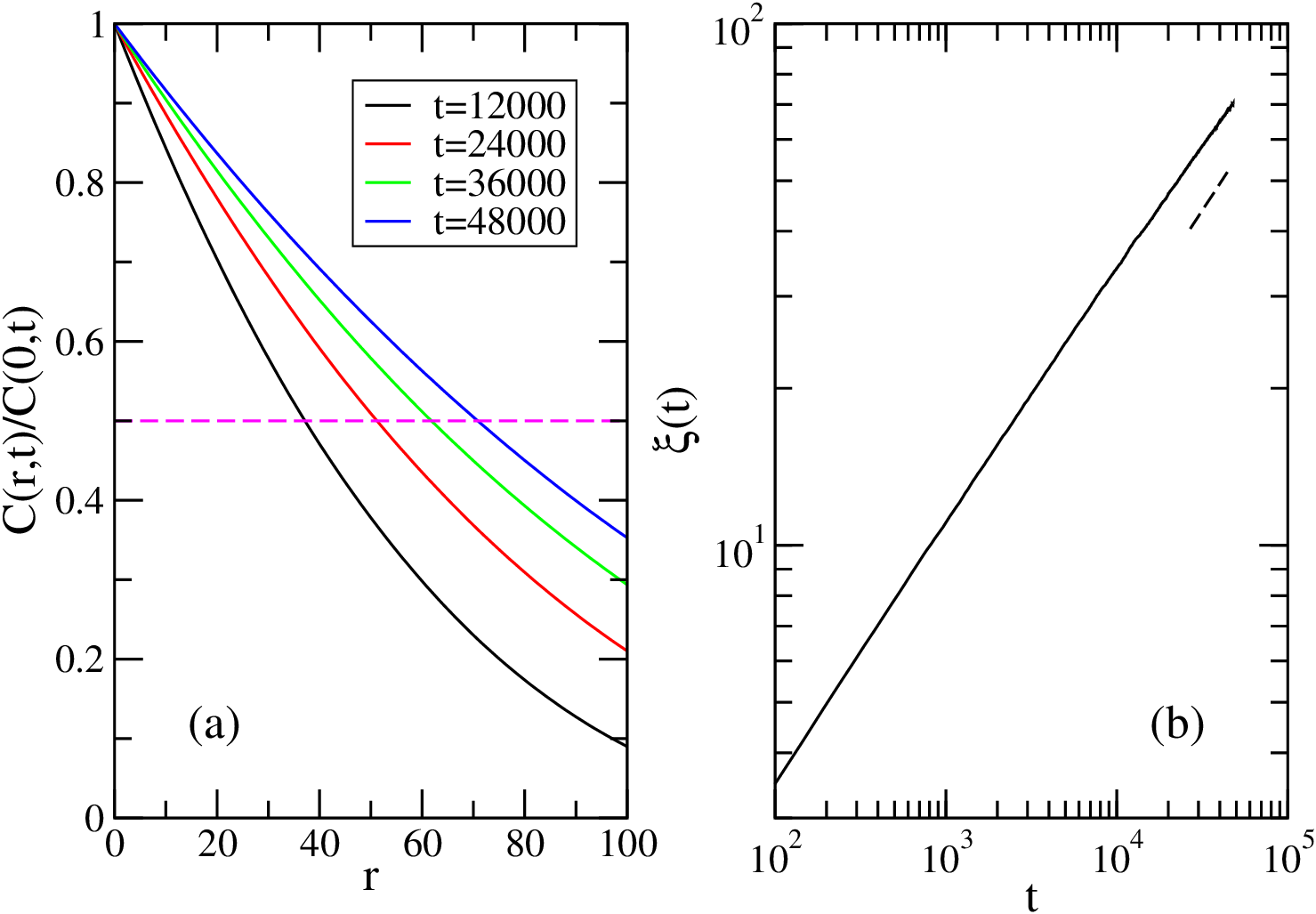}
\caption{\label{fig1} 
(a) Normalized space-time height-height correlation 
	function $C(r,t)/C(0,t)$ for free flux lines with $L = 2560$ as a 
	function of distance $r$ for four different times. The intersections 
	with the dashed magenta line $C_0 = 0.5$ yield a time-dependent 
	characteristic length. 
	(b) Time-dependent correlation length $\xi(t)$ obtained from $C(r,t)$. 
	The dashed line indicates the expected slope $1/2$ for free diffusive 
	motion.}
\end{figure}

Fig.~\ref{fig1} shows the space-time height-height correlation function (a) and 
the derived characteristic length $\xi(t)$ (b) for the case of free lines 
({\em i.e.}, non-interacting vortices in the absence of any disorder). As 
expected for this simple case, free lines undergo purely diffusive transverse 
motion, resulting in a square-root increase with time of the dynamic length 
scale, as indicated by the dashed line in Fig.~\ref{fig1}(b). The free-line 
data will serve as baseline in the next Section when we turn to the effects of 
the different contributions in the Hamiltonian on the growing length scale
$\xi(t)$.

\section{Results}

In the following discussion of our Monte Carlo simulation data, we first focus 
on samples with point-like defects, before considering later systems with
columnar pinning centers. As we will see, these different defect types reveal 
themselves through distinct characteristic features in the time-dependent 
correlation length.

\subsection{Point defects}

Point-like pinning sites are localized crystal defects, as for example oxygen 
vacancies in ceramic superconductors, that may occur naturally or can be 
introduced artificially. Irrespective of their origin, they exert a short-range
attractive force on the magnetic flux lines. These pinning sites are normally 
located at random positions in the sample. However, point-like defects can also 
be distributed in non-random ways; {\em e.g.,} recent intriguing studies have 
investigated the enhanced pinning efficiency of conformal pinning arrays 
\cite{Day13, Day14a, Day14b}.

\begin{figure}[h]
\includegraphics[width=0.80\columnwidth]{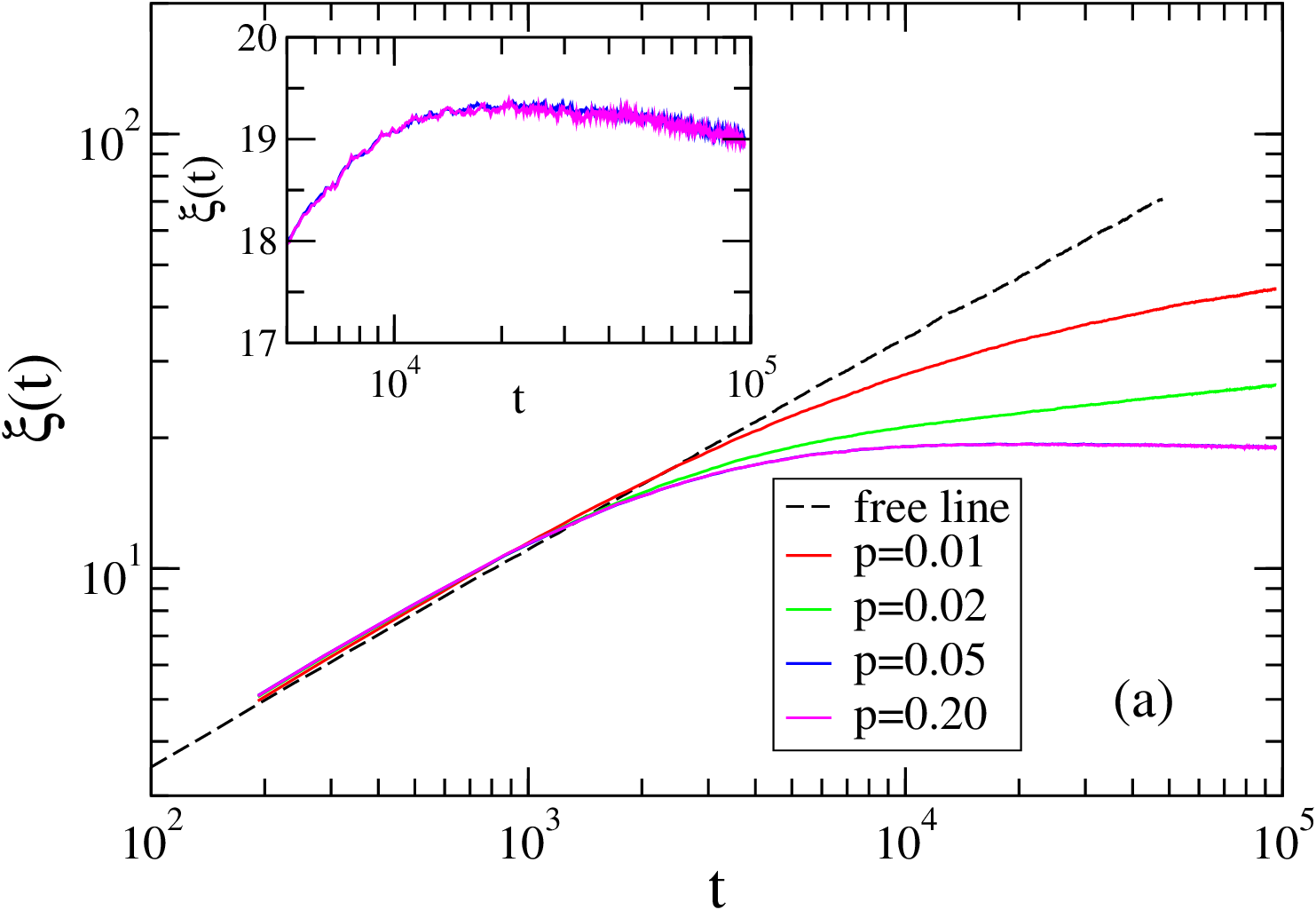}\\[0.3cm]
\includegraphics[width=0.80\columnwidth]{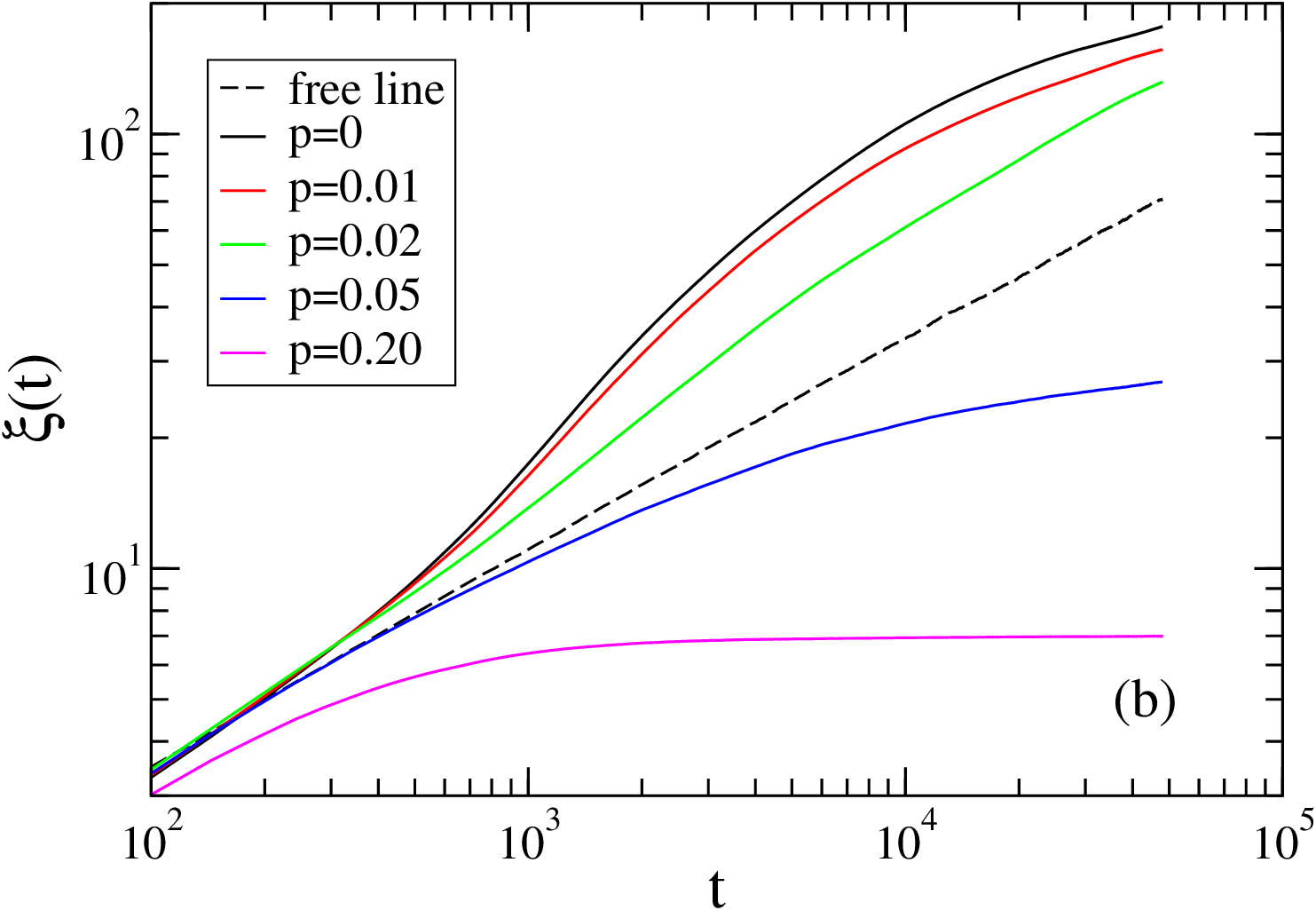}
\caption{\label{fig2} 
Correlation length $\xi(t)$ derived from the space-time 
	height-height correlation function (\ref{eq:C2}) in the presence of 
	point-like pinning centers for vortex lines of length $L = 2560$: 
	Figure~(a) shows the time evolution for non-interacting flux lines, 
	whereas the correlation lengths for interacting vortices are shown in 
	(b). The inset in (a) focuses on the long-time behavior of 
	non-interacting lines in the presence of strong pins. In both figures 
	the dashed line represents the correlation length of a free line, 
	{\em c.f.} Fig~\ref{fig1}(b).}
\end{figure}

Figure~\ref{fig2} summarizes our findings for randomly placed attractive 
pinning sites. First, we consider in Fig. \ref{fig2}(a) the simplified case of 
non-interacting flux lines in a random pinning landscape, before we turn to the
full problem of a thermal system of interacting magnetic vortices in a 
three-dimensional space with point pinning centers in Fig. \ref{fig2}(b). 

For non-interacting vortices in a disorder-free environment, the displacement
components of a flux line element in the plane perpendicular to the $z$ 
direction satisfy the Edwards-Wilkinson (or stochastic diffusion) equation. 
Consequently, the correlation length obtained from the space-time height-height 
correlation function for lateral line fluctuations (\ref{eq:C}) increases with 
time as $\xi(t) \sim t^{1/2}$, see Fig.~\ref{fig1}(b). The introduction of
pinning centers constrains the transverse vortex fluctuations, as some flux 
line elements are trapped at favorable near-by attractive pinning sites. This 
trapping can be temporary and transient for weak pinning strengths, where 
thermal noise is strong enough to allow the escape from the potential wells, or 
become permanent for strong pinning centers. These different situations are 
clearly reflected in a change of the behavior of $\xi(t)$ when increasing the 
pinning stength $p$, as demonstrated in Fig. \ref{fig2}(a). Inspection of the
graphs reveals that the correlation length grows diffusively at early times, 
just like that of a free line. The randomly placed straight vortices first need
to roughen thermally and then explore their disordered environment before a 
sizeable number of flux line elements reaches close enough to pinning sites to 
become trapped. Once they are captured, lateral flux line fluctuations are 
impeded, and hence the correlation length increase is slowed down when compared 
to the behavior of a free line. As one would expect, the crossover time at 
which deviations from the free-line behavior are observed decreases noticeably
with increasing pinning strengths $p$. For weak strengths, {\em c.f.} the data 
for $p = 0.01$ and $p = 0.02$ in Fig.~\ref{fig2}(a), the correlation length 
maintains an algebraic temporal behavior $\xi(t) \sim t^{1/z}$ in the long-time 
limit, yet with a dynamic scaling exponent $z > 2$. Thus for $p = 0.01$ we 
obtain the value $1/z \approx 0.15$, whereas for $p = 0.02$ we measure 
$1/z \approx 0.09$. For stronger pinning strengths, {\em e.g.}, $p = 0.05$ and 
$p = 0.20$, $\xi(t)$ becomes effectively independent of the value of $p$, and 
stops increasing after around $10^4$ Monte Carlo time steps. For such strong 
point pinning centers, the vortices are permanently trapped for the duration of
our simulation, and different flux line segments remain largely uncorrelated. 
Closer inspection of $\xi(t)$ for large $p$, depicted in the inset of 
Fig.~\ref{fig2}(a), reveals even a slight decrease of $\xi(t)$ for large $t$ as 
additional flux line elements are captured by the point defects, thus further 
reducing the lateral vortex fluctuations. We remark that we have observed 
qualitatively similar behavior both for the flux line mean-square displacement 
and radius of gyration in Langevin molecular dynamics simulations, see Fig.~5 
in Ref.~\cite{Dobramysl2013}.

Additional interesting features emerge in our simulations of the full problem 
of interacting vortex lines in a pinning landscape with attractive point 
defects. As shown in Fig.~\ref{fig2}(b), marked differences between weak and 
strong pins are again revealed by the correlation length $\xi(t)$. For weak 
pinning strengths three different regimes can be observed. At very early times 
the vortices still behave essentially like free lines, and the transverse
correlations along the lines are not yet affected by the mutual repulsive 
vortex interactions or the pinning to localized defect sites. This early-time 
regime, however, is very short, as it lasts only a few hundred Monte Carlo time
steps. After that the flux lines start rearranging themselves, build up 
longer-range spatial correlations, and tend to form a triangular lattice, which 
is revealed by enhanced values of $\xi(t)$ when compared to the characteristic
length scale of the free line. The full black line in Fig.~\ref{fig2}(b), which
refers to the situation without pins, clearly indicates the temporal changes of 
the correlation length when the Abrikosov flux lattice becomes established. 
Adding weak pins, as for example for $p = 0.01$ (red line in the figure) does 
not have a major impact on the correlations as the relatively strong 
vortex-vortex interactions prevent efficient pinning of flux line elements. 
Upon further increasing the pinning strength, the flux line interactions lose 
against the attractive point disorder, and more and more flux line elements get 
trapped. This induces a slowing down of the growth of $\xi(t)$ and, ultimately, 
a termination of the further increase of this characteristic length. Note that 
for the largest values of $p$ employed in our simulations, the ultimate 
correlation length for interacting vortices is much smaller than for 
non-interacting flux lines, compare the case $p = 0.20$ in Fig.~\ref{fig2}(a) 
and Fig.~\ref{fig2}(b). In fact, flux lines are quickly pushed away from their 
random initial locations due to their repulsion from the other vortices, and 
thus faster reach favorable pinning centers from which they cannot escape within
the simulation time window. We again note that similar temporal ranges are
observed for strong point defects in Langevin molecular dynamics simulation 
data for the mean-square displacement and gyration radius of mutually repelling
flux lines, {\em c.f.} Fig.~9 in Ref.~\cite{Dobramysl2013}; yet it is also
important to realize that the detailed crossover times separating the distinct
dynamical regimes do depend on the precise observable under investigation.

\subsection{Columnar defects}

As for point-like disorder, correlated columnar defects can appear naturally in 
a superconductor sample or can result from tailored manipulations of the sample.
Line dislocations provide an example of naturally occurring linearly extended 
defects of this type. High-energy ion radiation yields material damage tracks 
used to pin vortex lines in type-II superconductors; alternatively, linear
pinning centers, even with pre-arranged spatial distribution, can be engineered
during material growth. Columnar defects are known to provide a much more 
efficient pinning mechanism for flux lines than uncorrelated point disorder 
\cite{Civ91,Blatter_etal:94}.

In our comparative study of point-like and columnar pinning centers, we 
distribute the same number of pinning sites per layer; the crucial difference 
is that for columnar defects this pattern repeats itself from layer to layer.
It follows that in the presence of columnar pins, flux lines are much less 
likely to encounter a randomly placed defect. One therefore expects an average 
behavior much closer to that of vortices moving in a disorder-free environment.
Yet on the other hand, once a vortex segment is captured by an extended pinning
site that is correlated across the sample layers, there is a high probability 
that the whole flux line will become trapped \cite{Dobramysl2013}. Such 
vortices that are localized at a linear defect along their entire length 
hardly ever escape this extended pinning potential.

\begin{figure} [h]
\includegraphics[width=0.80\columnwidth]{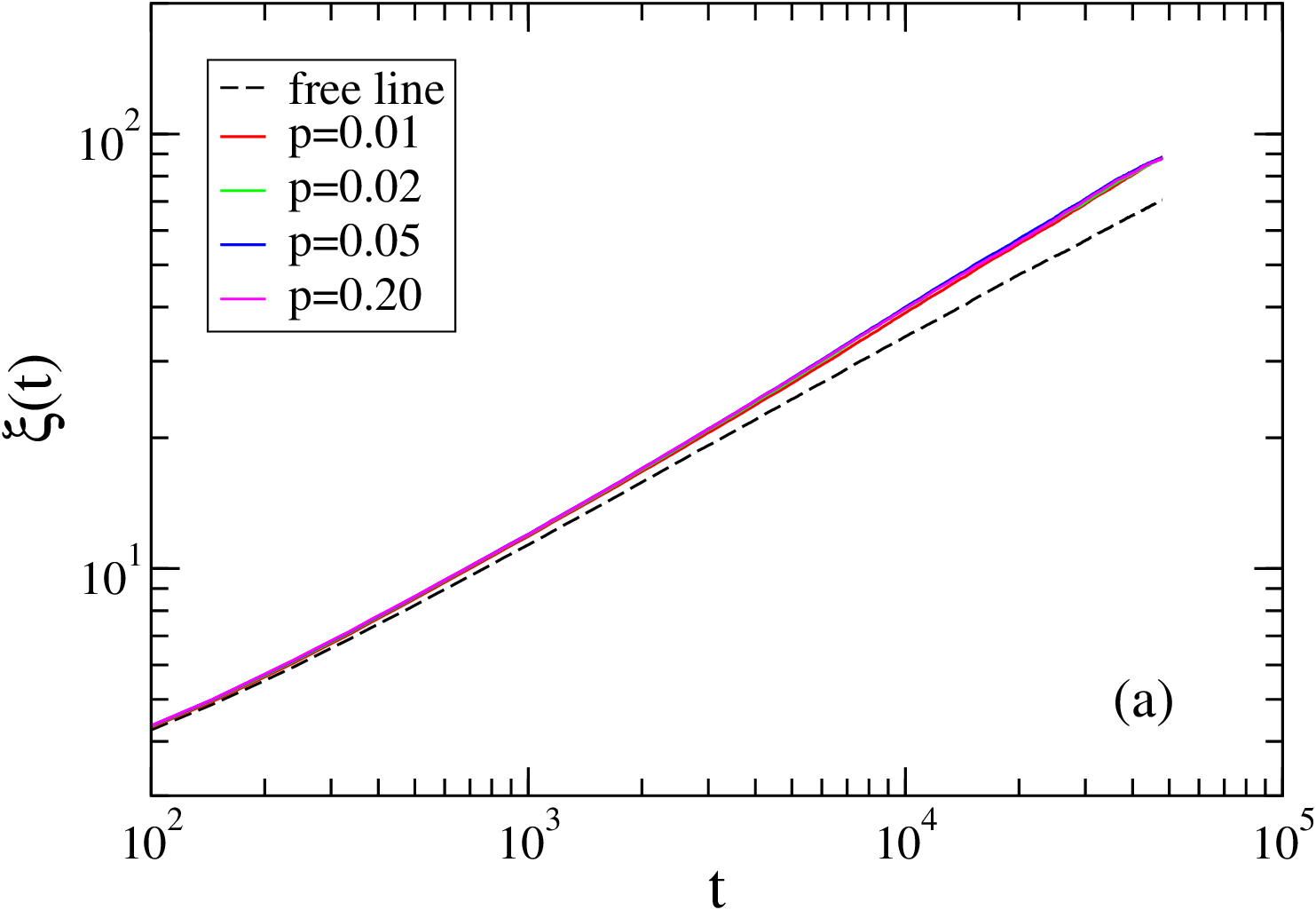}\\[0.3cm]
\includegraphics[width=0.80\columnwidth]{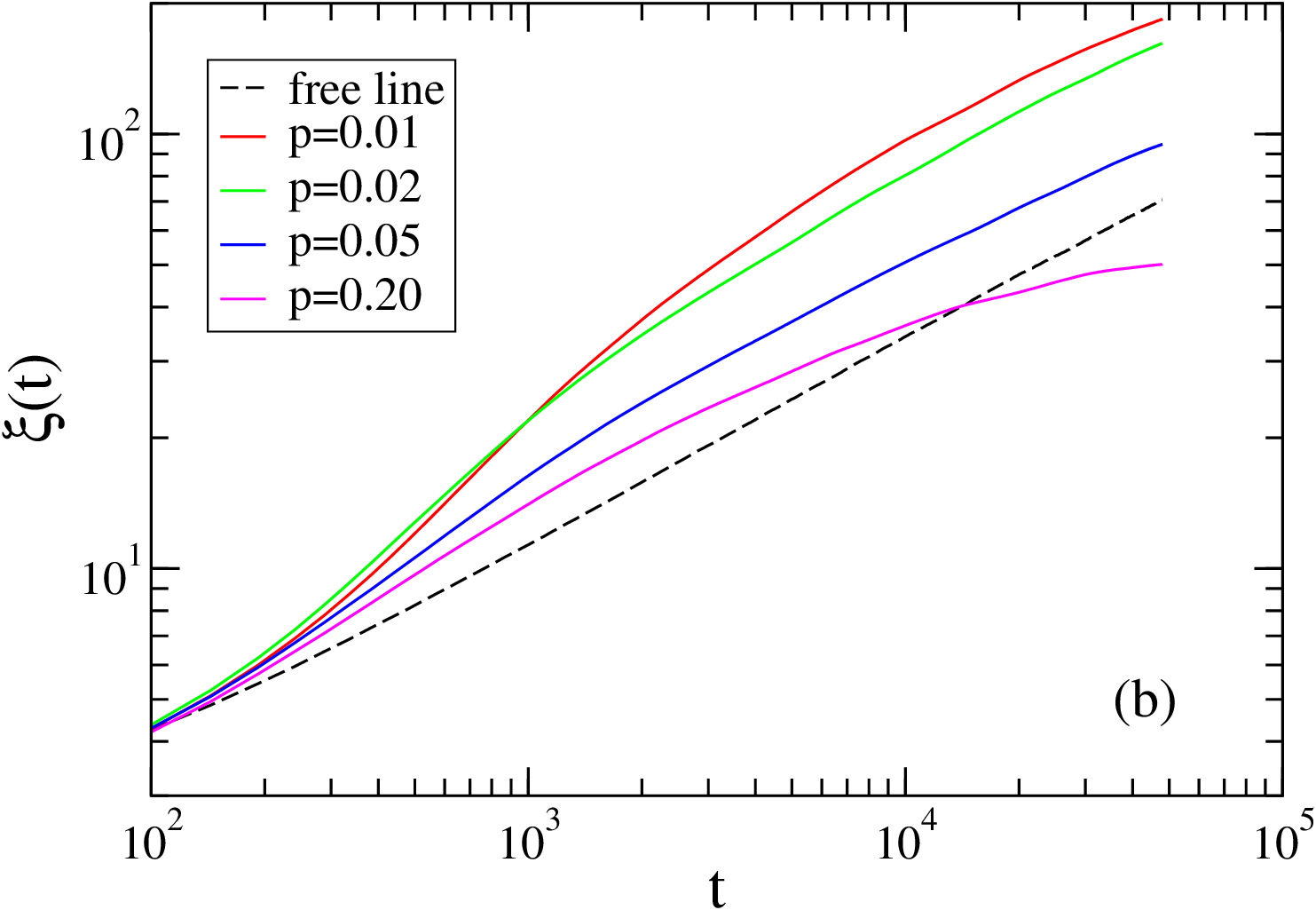}
\caption{\label{fig3} 
Time-dependent correlation length $\xi(t)$ in the 
	presence of columnar defects. Figure~(a) shows the time evolution for 
	non-interacting vortices, whereas in (b) the correlation lengths for 
	interacting flux lines are shown. In both figures the dashed line 
	represents the correlation length of a free line. The vortex lines are 
	composed of $L = 2560$ elements.}
\end{figure}

The time-dependent correlation length displayed in Fig.~\ref{fig3} reflects 
both the behavior of a free vortex at early times and the pinning of whole 
lines to columnar defects at later times. Let us first focus on the case of 
non-interacting vortices shown in Fig.~\ref{fig3}(a). We observe that in the
absence of mutual repulsion between the flux lines, the behavior of the 
characteristic length $\xi(t)$ is independent of the pinning strength. This 
apparent universal feature is of course due to the fact that a vortex captured 
by an extended columnar pin cannot escape from this trap but instead remains 
firmly localized at the defect. We also note that the correlation length in a 
system with extended defects only displays rather small deviations from the 
free-line behavior. This is readily understood by remembering that the features
revealed in $\xi(t)$ result from superposition of the height-height correlations
of lines that have remained free, {\em i.e.}, not captured by defects, and of 
vortices trapped by columnar pins whose transverse fluctuations are much 
suppressed. At first look it might be puzzling that the correlation length in 
systems with defect lines is slightly {\it larger} than for the free vortices. 
This behavior is explained by the observation that different segments of very 
long flux lines can be caught by spatially separated columnar defects. These 
captured parts are connected by segments with high elastic energy and strong 
correlations between the flux line elements. In order to check this 
interpretation we also ran simulations for much shorter vortex lines (comparable
to those studied in Ref.~\cite{Dobramysl2013}) where it is unlikely that the 
vortices are caught by more than one line defect. As expected, the correlation 
length for these smaller lines is always below that measured for free lines.

The situation becomes considerably more complicated for the full problem of 
interacting vortex lines subject to columnar pinning centers. We infer from 
Fig.~\ref{fig3}(b) for small pinning strengths $p$ a behavior for $\xi(t)$ that
is qualitatively similar to that encountered for point-like defects: Due to 
their mutual repulsion, the vortices are displaced from their initial positions
and tend to form an Abrikosov lattice, a process only slightly impeded by the 
randomly placed columnar defects in the system. For stronger defects, one 
essentially observes again the superposition and mixing of two different types 
of behavior, namely that of hitherto unbound lines that try to arrange 
themselves at least locally in a triangular lattice, and the features of lines 
that have already become trapped by defect columns and display reduced lateral
fluctuations. However, in contrast to the data for non-interacting lines shown 
in Fig.~\ref{fig3}(a), these two subsets of flux lines are not completely 
independent, since their dynamics and correlations become connected through the 
vortex-vortex interactions. As a result, the correlation length $\xi(t)$ keeps 
increasing with time, even for pinning strengths for which in samples with 
point-like defects we observed rapid saturation, compare Fig.~\ref{fig2}(b) 
with Fig.~\ref{fig3}(b) for the case $p = 0.20$. This behavior differs from 
that of the gyration radius that changes non-monotonically with time, but is 
similar to the time-dependence of the flux line mean-square displacement
\cite{Dobramysl2013}.

\section{Conclusion}

Vortices in type-II superconductors display very complex non-equilibrium 
properties, owing to the competition between different but comparable energy 
scales: elastic flux line tension, mutual vortex repulsion, defect pinning, and 
thermal noise. Using extensive Metropolis Monte Carlo simulations for an 
effective elastic line model with realistic parameters for YBCO, we aimed at 
investigating the non-equilibrium relaxation processes in this system through 
extracting a time-dependent correlation length $\xi(t)$ from the space-time 
lateral flux line height-height correlation function. We found that this 
characteristic length displays very rich behavior that allows us to identify 
different temporal relaxation regimes. Moreover we observe markedly different
time dependences for point-like and extended columnar attractive defects that 
originate from the quite distinct underlying fluctuation spectra.

In previous studies of the relaxation kinetics of vortex matter in disordered
type-II superconductors \cite{BCD1, BCD2, BCI, IBKC, Pleimling2011,
Dobramysl2013, Assi2015}, various single-time quantities and two-time 
correlation functions, for example the two-time height-height autocorrelation 
function, the two-time mean-square displacement, and the two-time vortex 
density-density autocorrelation function, have been measured. Two-time 
quantities often yield important insights into relaxation processes far from 
equilibrium, but the physical interpretation of their behavior can also be 
quite challenging. The correlation length studied in this work has the virtue 
of being simple and its signatures comparatively easy to understand. In 
addition, its characteristic features yield insights into different temporal 
relaxation regimes and into the emerging distinctions between uncorrelated
point pins and extended linear defects.

Time-dependent correlation lengths have been the focus of several recent 
numerical investigations that probed non-equilibrium relaxation processes in 
disordered systems \cite{Par10, Cor11, Cor12, Par12, Cor13, Man14}. 
However, most of these studies dealt with rather simple models as, {\em e.g.},
disordered Ising systems. Our work reveals that a time-dependent length scale
can also be extracted for much more complex situations exemplified here by 
vortex matter subject to thermal fluctuations, disorder, and long-range
interactions. In disordered ferromagnets, the relevant growing length scale 
$L(t)$ obtained from space-time correlations does not obey a simple power law
either. Hence it is crucial in the non-equilibrium aging scaling regime to
plot the data for associated two-time autocorrelations not simply as functions
of the time ratio $t / s$, but instead against $L(t) / L(s)$ in order to attain
convincing data collapse \cite{Par10, Par12}. It is tempting to try a similar
approach in the considerably more complex disordered vortex matter system under
investigation here, for which previous Monte Carlo as well as Langevin 
molecular dynamics simulations have yielded very complicated behavior in their
non-equilibrium relaxation kinetics \cite{Pleimling2011, Dobramysl2013, 
Assi2015}. We have indeed checked if data scaling collapse ensues if we plot
our Monte Carlo aging data for two-time correlation functions of 
Ref.~\cite{Pleimling2011} versus $\xi(t) / \xi(s)$ with the characteristic 
length scales measured in this current work. However, the results turned out
unsatisfactory; in fact, the various competing energy contributions to the
Hamiltonian (\ref{hamilt}) induce intricate crossovers between distinct 
relaxation time regimes, which however differ depending on the observable under
consideration. Hence there does not exist a single characteristic and dominant
length scale that governs all physical quantities of this complex system, which
renders any truly simple aging scaling picture obsolete.

\ack
This research is supported by the U.S. Department of Energy, Office of Basic
Energy Sciences, Division of Materials Sciences and Engineering under Award
DE-FG02-09ER46613.

\end{document}